\documentclass{article}

\usepackage{arxiv}

\usepackage[utf8]{inputenc} 
\usepackage[T1]{fontenc}    
\usepackage{hyperref}       
\usepackage{url}            
\usepackage{booktabs}       
\usepackage{amsfonts}       
\usepackage{nicefrac}       
\usepackage{microtype}      
\usepackage{lipsum}
\usepackage{graphicx}
\usepackage[english]{babel}
\graphicspath{ {./images/} }

\title{Composition-property extrapolation for compositionally complex solid solutions based on word embeddings}

\author{
Lei Zhang\\
Interdisciplinary Centre for Advanced Materials Simulation\\
Ruhr-University Bochum\\
Universit\"atsstra\ss e 150\\
44780 Bochum, Germany\\
\texttt{lei.zhang-w2i@rub.de}\\
   \And
Lars Banko\\
Chair for Materials Discovery and Interfaces\\
Institute for Materials\\
Ruhr-University Bochum,
Universit\"atsstra\ss e 150\\
44780 Bochum, Germany\\
	\And
Wolfgang Schuhmann\\
Analytical Chemistry - Center for Electrochemical Sciences (CES)\\
Faculty of Chemistry and Biochemistry\\
Ruhr-University Bochum\\
Universit\"atsstra\ss e 150\\
44780 Bochum, Germany
	\And Alfred Ludwig\\
Chair for Materials Discovery and Interfaces\\
Institute for Materials\\
Ruhr-University Bochum,
Universit\"atsstra\ss e 150\\
44780 Bochum, Germany\\
	\And
	Markus Stricker\\
Interdisciplinary Centre for Advanced Materials Simulation\\
Ruhr-University Bochum\\
Universit\"atsstra\ss e 150\\
44780 Bochum, Germany\\
\texttt{markus.stricker@rub.de}
}

\begin{document}
\maketitle
\begin{abstract}
Mastering the challenge of predicting properties of unknown materials with multiple principal elements (high entropy alloys/compositionally complex solid solutions) is crucial for the speedup in materials discovery. We show and discuss three models, using property data from two ternary systems (Ag-Pd-Ru; Ag-Pd-Pt), to predict material performance in the shared quaternary system (Ag-Pd-Pt-Ru).
First, we apply Gaussian Process Regression (GPR) based on composition, which includes both Ag and Pd, achieving an initial correlation coefficient for the prediction ($r$) of 0.63 and a determination coefficient ($r^2$) of 0.08.
Second, we present a version of the GPR model using word embedding-derived materials vectors as representations.
Using materials-specific embedding vectors significantly improves the predictive capability, evident from an improved $r^2$ of 0.65.
The third model is based on a `standard vector method' which synthesizes weighted vector representations of material properties, then creating a reference vector that results in a very good correlation with the quaternary system's material performance (resulting $r$ of 0.89). Our approach demonstrates that existing experimental data combined with latent knowledge of word embedding-based representations of materials can be used effectively for materials discovery where data is typically sparse.
\end{abstract}


\section{Introduction}
Material science is a driver of technological progress by development of innovative materials that enable advancements across industries from electronics to aerospace~\cite{Wang2012699,Suryanarayana20011}.
Novel materials are the driver because of new properties or property combinations or by replacing existing critical or expensive materials with less critical ones while at the same time not sacrificing performance.
Discovering new materials (fast) requires to accurately predict material properties, particularly in complex materials such as compositionally complex solid solutions (``high entropy alloys'') with four or more primary elements.
Such systems show promise as \textit{Discovery Platforms}, e.g. for electrocatalysis~\cite{Batchelor2019}.
However, they pose significant challenges for discovery since the possible combinations of elements and their compositional ratios renders brute-force screening approaches practically impossible.
Additionally, predicting their properties is difficult due to their complex compositional interactions and the intricate ways in which these interactions affect material behavior~\cite{Leshchenko2018}.
As such, the acceleration of the discovery process for new materials necessitates the development of new methods to navigate complex composition-structure-property relationships of promising material systems.

The integration of computational power and data analysis is necessary in overcoming the challenges presented by these material systems~\cite{Back2024}.
Machine learning has emerged as a useful tool, providing a path for material scientists to predict and understand the properties of materials system~\cite{Zhao2022,Persaud2024}.
This transition from traditional, heuristic approaches to data-driven, computational strategies signifies a transformation of the field\cite{Durdy2022,Lu2022}, aligning with the complexity of the possible materials of interest.
Among data-centered approaches, Gaussian Process Regression (GPR) has demonstrated exceptional versatility and efficacy across multiple domains, illustrating its capacity to model complex relationships~\cite{JOHANSEN1991}.
The adaptability of GPR stems from its non-parametric approach which allows to adjust its complexity based on the dataset, a feature that sets it apart from models like neural networks~\cite{Lu2021}.
This flexibility renders it particularly valuable in applications for complex non-linear relationships in high-dimensional data spaces.
However, the usefulness, that is the predictive power, hinges on available data and meaningful representations.
Often, sophisticated adjustments to such models are necessary to effectively capture complex correlations~\cite{Menke2021}.
Possible modifications include appropriate accounting for noise in data and customization of kernel functions.
A critical part of any data-based approach is the representation\cite{Einarsson2020293,Bayerlein2022,Damewood2023399} of the input.
In particular, the challenge of how to represent \textit{a material}. 
A simple approach is to just use the composition~\cite{Thelen2023}.
This is often sufficient for interpolation. 
However, if the goals is to predict into unknown spaces, any existing knowledge about a material or similar materials and their properties is desirable.
In this, the vast expanse of scientific literature represents a rich, yet not fully exploited, resource~\cite{Jablonka2023}.
Through literature mining\cite{Weston20193692,Gupta2022} and vector analysis\cite{Ward201860}, we can convert the latent knowledge contained in scientific texts into formats amenable to machine learning in form of representations~\cite{Subramanian2023}.
The integration of word embedding-based vector analysis, derived from literature mining, with machine learning models like GPs, represents a new path for improving predictive capabilities in for materials discovery, particularly for complex systems such as ternary and quaternary materials.\\
In our example, we present the problem of predicting the performance of a quaternary system for electrochemical applications, specifically the oxygen reduction reaction (ORR). Here, ``performance'' is defined as the current density of electrocatalysis of the ORR at an overpotential of 850 mV.
We use existing measurements of ternary systems in conjunction with representations of materials and properties based on word embeddings.
We examine three distinct approaches: \textit{standard} GP modeling based on composition, GP augmented with material vectors based on word embeddings, and our `standard vector method'.

Our approach enhances the prediction capabilities into compositionally more complex materials by combining measured data from compositionally less complex materials, combined with advanced representations of materials through word embeddings. We illustrate its predictive power and compare it with the reference approach that solely relies on materials representations based on composition.

\section{Methods}

\subsection{Dataset Description}
For our demonstration we use two datasets from two different overlapping ternary systems (Ag-Pd-Ru and Ag-Pd-Pt) to train models for property prediction of a shared quaternary system (Ag-Pd-Pt-Ru).
The basic idea is to use compositionally less complex systems (ternary materials) to predict the behavior of more complex ones (quarternary) in the context of electrocatalysis, specifically the ORR~\cite{Dathar2012,Pedersen2021,Bampos2023}.
Two ternary datasets are used to fit models that capture their correlation with electrocatalytic properties, specifically a current at a fixed applied overpotential.
These models are then used to predict the electrocatalytic properties of the shared quaternary system, which includes all the elements present in the ternary systems.
The experimental data is sourced from composition-spread materials libraries which is described in detail elsewhere~\cite{Clausen2023}. Nevertheless, we provide a brief description here for completeness.
The materials libraries were fabricated by co-sputtering thin films on 100 mm diameter sapphire wafers (c-plane) from 4 elemental targets. The targets were confocally aligned to a 100 mm substrate (target-substrate distance approx. 12 cm). Target materials had a purity of 99.99,\%. Ar (99.9999\,\%) was used as a sputter gas. The deposition pressure was 0.667\,Pa. The film thickness was 100 - 150\,nm.
The chemical composition of the materials libraries was measured by energy dispersive X-ray spectroscopy (EDX) with an acceleration voltage of 20\,kV. 81 measurements were done on a regular grid of 9x9 (8.5\,mm spacing) on each library. Linear regression was used to interpolate the composition over the 342 measurement areas of a 4.5\,mm grid that was used in the scanning droplet cell experiments. 
Electrochemical measurements were conducted with the use of a high-throughput scanning droplet cell (SDC). The SDC head incorporates counter (Pt wire) and reference (Ag|AgCl|3\,M KCl) electrodes and a teflon tip with 1\,mm diameter. The materials library is connected as working electrode, e.g. the surface of the investigated sample in every spot where the tip touches the sample. The electrolyte was replaced for every measurement area. Linear sweep voltammograms were measured in 0.05\,M KOH, pH 12.5, with a scan rate of 10\,mV/s.
All potentials are reported versus the RHE according to the following equation:
URHE (V) = U(Ag|AgCl|3\,M KCl) + 0.210 + (0.059 pH),
where U(Ag|AgCl|3\,M KCl) is the potential measured versus Ag|AgCl|3\,M KCl reference electrode, 0.210\,V is the standard potential of the Ag|AgCl|3\,M KCl reference electrode at 25$^\circ$C. Note that 0.059 is the result of $(RT)\cdot(nF)^{-1}$, where R is the gas constant, T is the temperature (298\,K), F is the Faraday constant, and n is the number of electrons transferred during the reaction.

\subsection{Modeling Approaches}

\subsubsection{Method 1: Gaussian Process (GP) Model with Elemental Composition}
A Gaussian Process (GP) model based of elemental composition derived from the ternary datasets is fit to predict the electrochemical current at a potential of 850\,mV for the quaternary system. This sort-of traditional approach offers a robust framework for predictions about electrocatalytic performance.
In materials science, GP models have been effectively applied to predict various properties, including thermal conductivity\cite{Zhang2020}and electronic structure~\cite{Wang2022}.
This model serves as our baseline, allowing us to evaluate the models with more nuanced representations against a reference standard.

\subsubsection{Method 2: Enhanced GP Model with Material Vectors}
The second model is different to the standard GP model by employing `material vectors' instead of the elemental composition as a representation for materials.
Material vectors are obtained from a Word2Vec model based on a comprehensive literature review~\cite{ZHANG2024101654}.
We retrieve a 200-dimensional vector representation of each pure element.
Within this 200-dimensional space, we create representations for materials by a weighted linear combination of the elemental representations, in line with vector operations in word embedding space~\cite{Tshitoyan201995}.
By employing material vectors, we use the latent knowledge from scientific literature and transform it into an explicit, quantifiable form, to improve our model's predictive power.
Like our baseline GP model, we predict each material's electrocatalytic performance, enabling a direct comparison between these two approaches.

\subsubsection{Method 3: Standard Vector Method}
\begin{figure*}
    \centering
    \includegraphics[width=\linewidth]{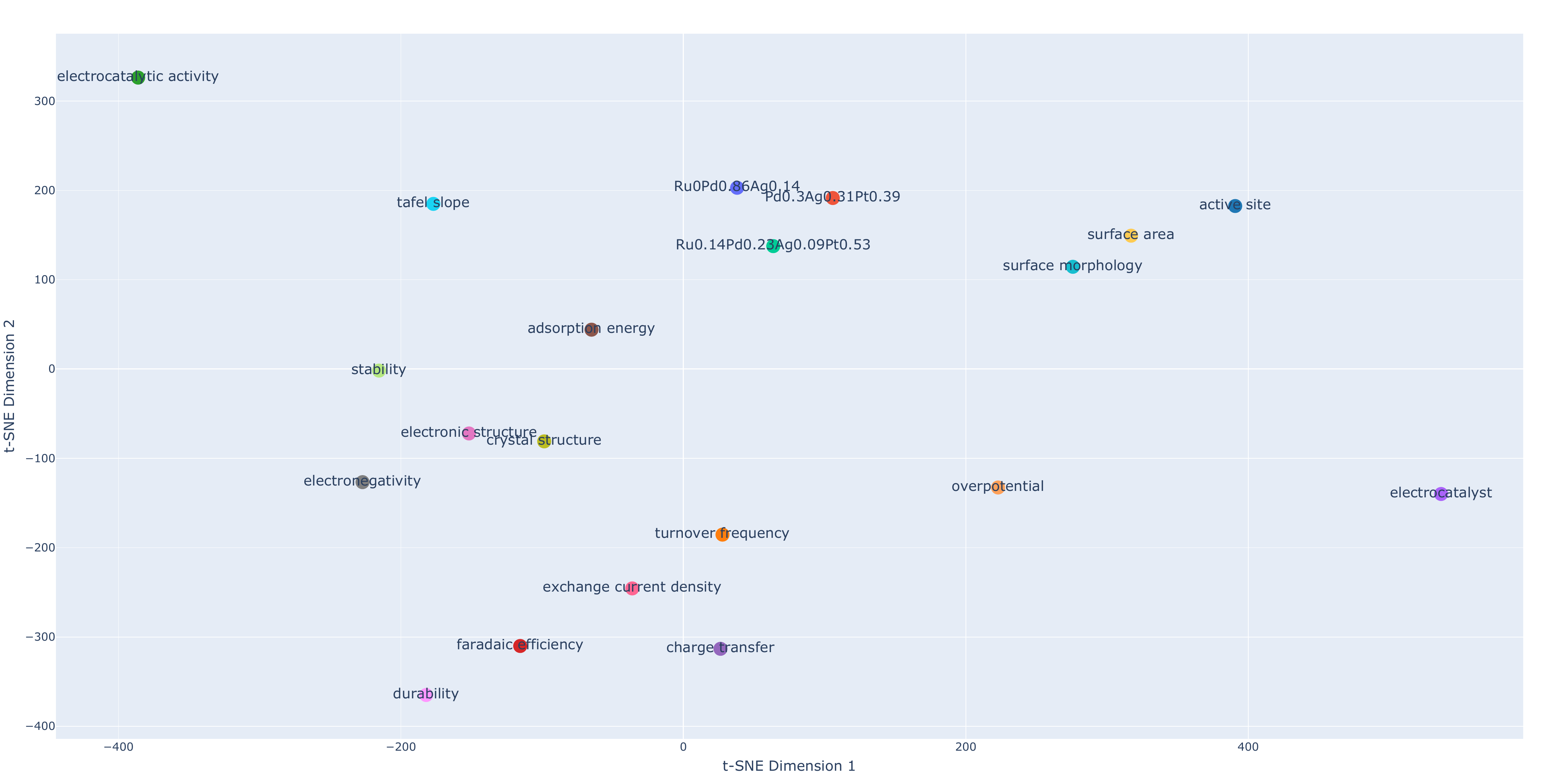}
    \caption{Dimensionality reduced (t-SNE) map of vector representations for the chosen electrocatalytic properties and pure elements.}
    \label{fig:properties_materials_vectors}
\end{figure*}

The third method is different from the GP-based models in two aspects.
For one, we introduce a novel approach based on the concept of a `standard vector'.
The idea is to substitute representations of compositions based on word embeddings of elements and their linear combinations with a similarity vector obtained by comparing with known terms related to electrocatalysis, thereby encoding explicit domain knowledge in the representation of a material.
The similarity of each word embedding representation of the composition with the term constitute \textit{one dimension} of the standard vector.
The process begins with the assembly of a list of material properties relevant to electrocatalysis, from which vector representations are generated.
Our property list include ``electrocatalyst'', ``overpotential'', ``tafel slope'', ``exchange current density'', ``stability'', ``durability'', ``surface area'', ``active site'', ``turnover frequency'', ``electrocatalytic activity'', ``faradaic efficiency'', ``charge transfer'', ``adsorption energy'', ``electronic structure'', ``electronegativity'', ``crystal structure'', and ``surface morphology'' -- a 17-dimensional space.
Fig.~\ref{fig:properties_materials_vectors} shows a dimensionality-reduced map of the vector representations for both, the listed terms, and the pure elements using t-SNE~\cite{Maaten2009}.

Each property in the list is chosen based on its known relationship to electrocatalytic performance and its role in determining the efficiency of the ORR. For instance, properties such as overpotential, tafel slope, and exchange current density are critical for assessing the electrocatalytic performance of materials.
Stability, durability, and surface area affect the longevity and effectiveness of catalysts in practical applications.
Other properties like adsorption energy, electronic structure, and crystal structure offer deeper insights into the interaction mechanisms at the molecular level which might influence catalytic behavior and performance.
The relative distance of word embeddings of materials to these properties capture the co-occurrence, and therefore proximity in embedding space.
Our hypothesis is that proximity of properties and materials representations in embedding space captures correlations and thereby provides an improved representation of materials, not based on their composition, but based on their latent properties and their relationships.

However, the novelty of our approach is in how these property vectors are combined.
Instead of simply merging the individual 17 similarity values, we calculate a `standard vector' that represents an ideal electrocatalyst by weighting each property vector based on the experimental data for the two ternary systems to reflect its importance w.r.t. known catalytic activity in this material system.
In essence, we create a \textit{reference} vector based on measured data which represents optimal characteristics for ORR performance for the given system.
The weighting step, a fitting procedure, is a minimization with constraints. The weights are adjusted to minimize the squared difference between `experimental indicators' (current at potential) and similarity dimensions.
In our case, we use measured activity as experimental indicator, but any reliable known data for materials correlating with the predicted property could be used in general.
We then assume that materials which are `closer' in vector space to this standard vector -- measured by similarity metrics such as cosine similarity -- are more likely to exhibit \textit{good} electrocatalytic performance.
By evaluating materials based on their proximity to this `ideal' vector, we predict and identify promising electrocatalysts without relying solely on only compositional or structural data.
Once defined, the standard vector based on the two ternary systems is a benchmark representation for evaluating materials in the shared quaternary system.
Rather than predicting performance by predicting the (measured) current directly, we apply similarity measures to pinpoint materials that align closely with the ideal standard vector, thereby identifying candidates with potentially high electrocatalytic performance.

\subsection{Statistical Analysis and Model Evaluation}
The performance of the first and second GP model is quantitatively assessed using the correlation coefficient ($r$) between the actual and predicted current densities, alongside the coefficient of determination ($r^2$), to gauge the models' ability to capture variance in the actual measurements.
This dual-metric approach allows for a comprehensive evaluation of the models' predictive accuracy and reliability.
The third model, employing the standard vector method, is assessed differently.
Given the different nature of its output, we adapt our evaluation strategy, using the correlation coefficient between the actual current densities and our predictions, the similarity scores.
This metric reflects the model's performance in identifying materials with high electrocatalytic performance based on their conceptual proximity to the `ideal' electrocatalyst as defined by the standard vector.
To further underscore the models' applicability to high-performance electrocatalysts, we introduce a filtering criterion, focusing on data points where the current at 850 mV (\texttt{Current\_at\_850mV}) is below -0.2 mA/cm\(^2\).
This is designed to improve the models' ability to identify materials with significant electrocatalytic activity. By focusing on data points where the current at 850 mV (\texttt{Current\_at\_850mV}) indicates notable activity, we tailor our analysis to emphasize materials that, based on our dataset, stand out for their electrocatalytic performance. This method allows us to direct our model's focus and analytical efforts towards those candidates most likely to impact future electrocatalysts.
In other words, for materials displaying low activity, we are not interested in `how low'.

\subsection{Validation and Reproducibility}

MatNexus~\cite{ZHANG2024101654}, an open-source tool developed by us, underpins our data processing, analysis, and visualization workflows.
MatNexus supports the standardized handling of materials science data, ensuring the reproducibility of our findings through a workflow.
We use it for all parts of the analysis: from initial data preprocessing to feature extraction, structuring for word embedding model training, and the visualization of datasets and analysis results.
We also use it to create a word embedding model to generate material vectors, which are then used in conjunction with the GP model as well as in the standard vector method for predictive analysis.
MatNexus is used to conduct targeted literature queries, focusing on articles indexed in Scopus with keywords `electrocatalyst' and `high entropy alloy' published before the year 2024.
We restrict our search to Open Access (OA) articles, primarily due to the uncertain legal implications of using copyright-protected materials without explicit permissions.
This approach not only aligns with our commitment to open science but also ensures compliance with copyright laws. Furthermore, in building our word embedding model, we limit our analysis to the abstracts of these papers, not the full texts, balancing depth of analysis with the accessibility of data (See the Supplementary Bibliography document).
For details of the implementation of MatNexus and its functionality, refer to our MatNexus repository on PyPI (\url{https://pypi.org/project/matnexus/})~\cite{ZHANG2024101654}.
All relevant codes, experimental datasets, and model predictions publicly accessible via GitHub (\url{https://github.com/lab-mids/ccss_word_embedding_prediction}), ensuring that our research can be validated, replicated, or expanded upon by others.

\section{Results}
\subsection{Dataset overview}

\begin{figure}
    \centering
    \includegraphics[width=0.75\linewidth]{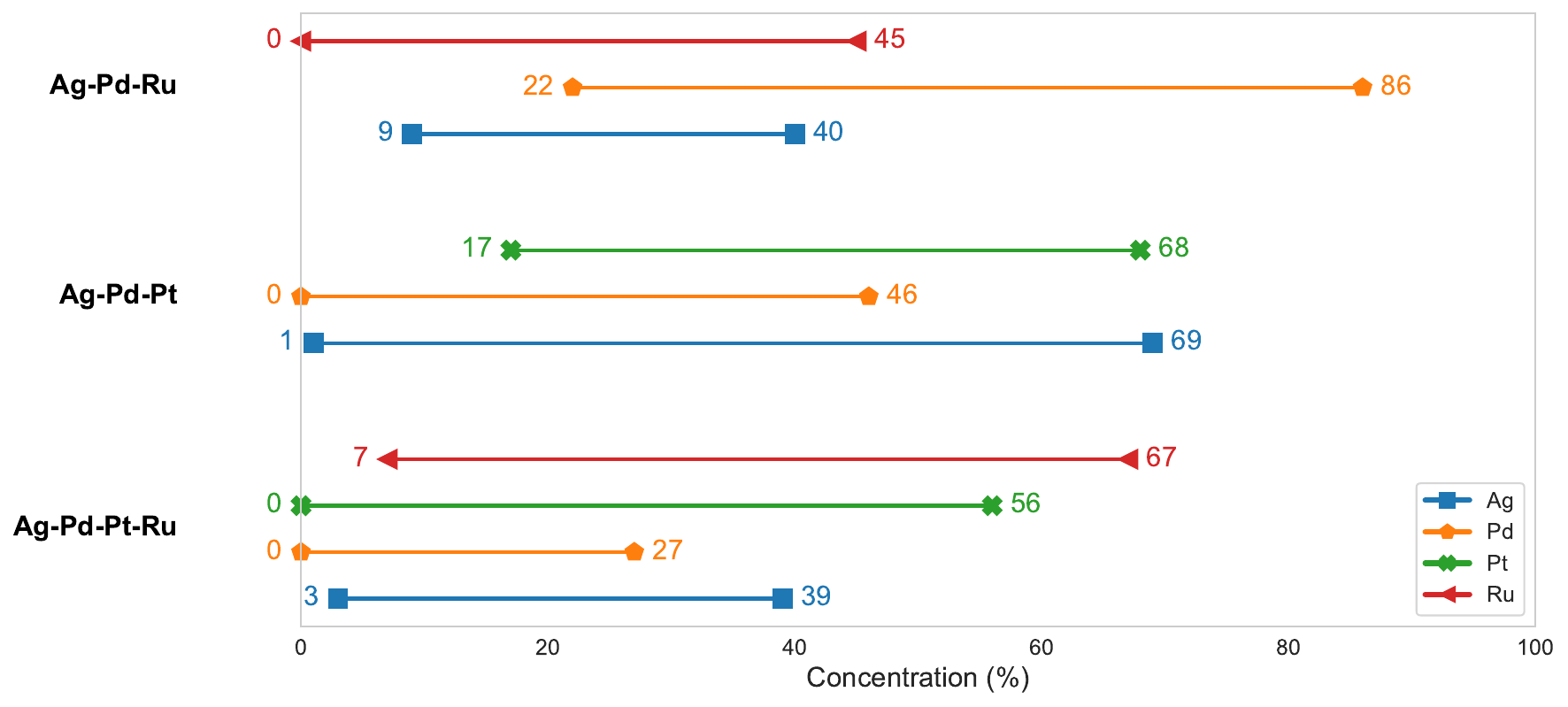}
    \caption{Compositional ranges of synthesised materials.}
    \label{fig:element_concentration_horizontal_range}
\end{figure}

\begin{figure}
    \centering
    \includegraphics[width=0.75\linewidth]{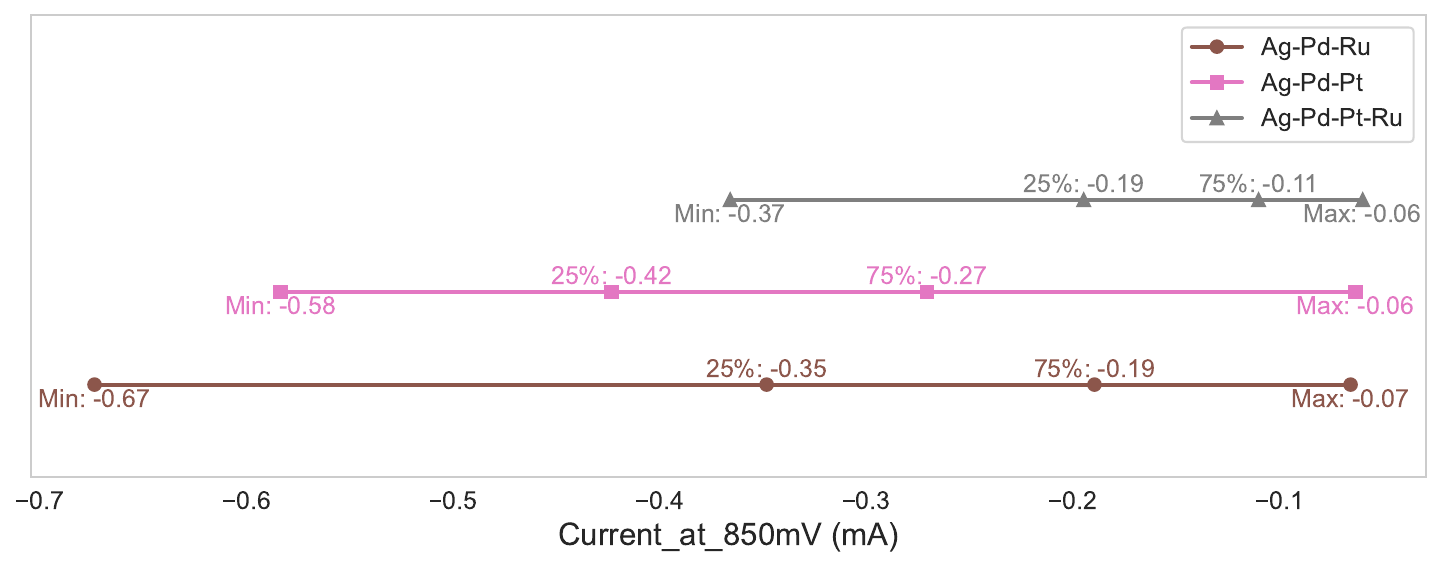}
    \caption{Current density ranges of synthesised materials.}
    \label{fig:current_density_horizontal_range}
\end{figure}

\begin{figure}
    \centering
    \includegraphics[width=0.5\linewidth]{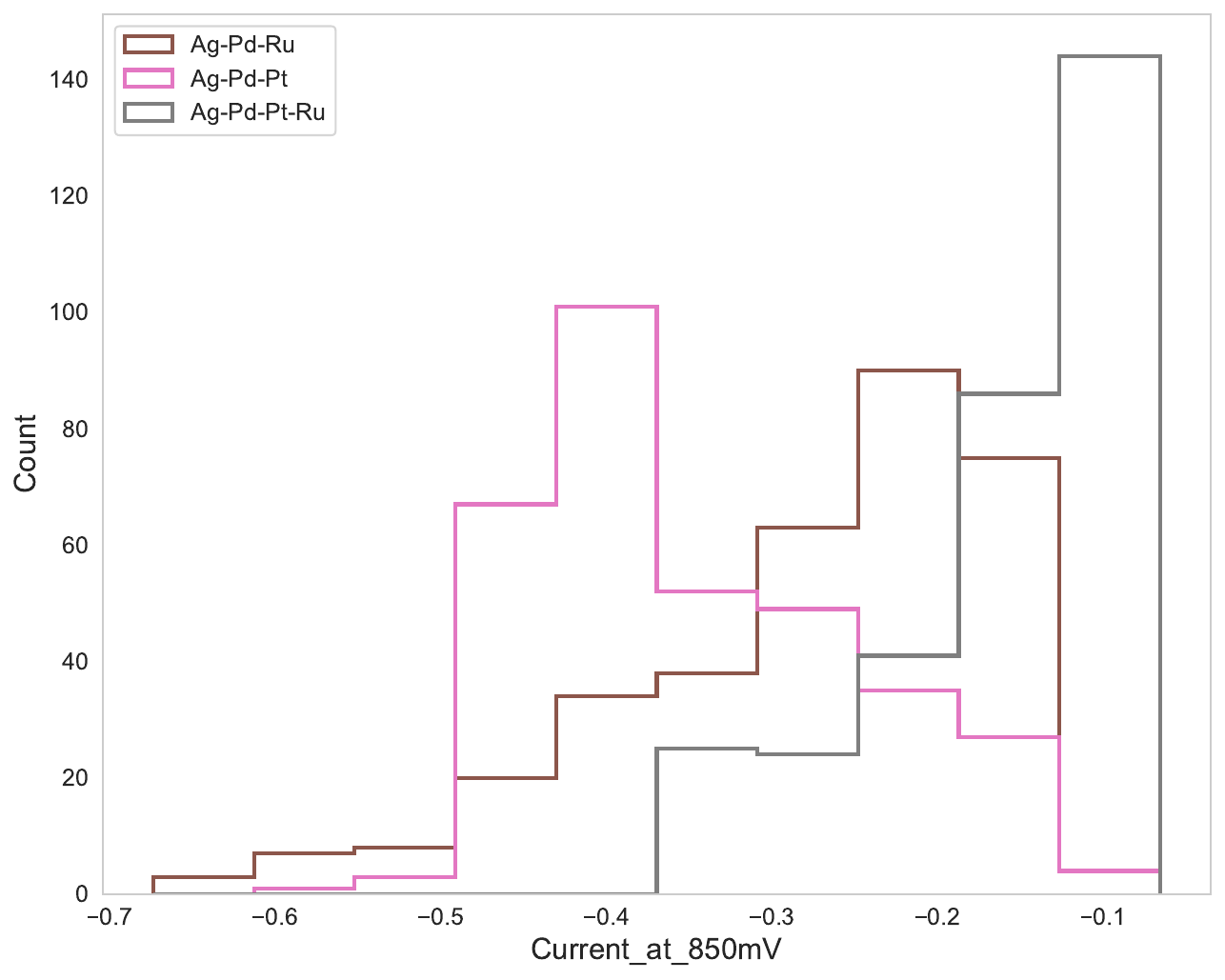}
    \caption{Stacked step histogram of current density across the samples.}
    \label{fig:current_density_stacked_step}
\end{figure}

This section provides an overview of the datasets used for model training and prediction (Table~\ref{tbl:elementalConcentrations}, Table~\ref{tbl:current850mVCorrelation}), Fig.~\ref{fig:element_concentration_horizontal_range}, Fig.~\ref{fig:current_density_horizontal_range}, Fig.~\ref{fig:current_density_stacked_step}).
The training datasets comprise two ternary systems (Ag-Pd-Ru;Ag-Pd-Pt), the prediction target data set is their shared quaternary system (Ag-Pd-Pt-Ru).

\begin{table*}
\small
\caption{Comparative elemental composition across systems.}
\label{tbl:elementalConcentrations}
\begin{tabular*}{\textwidth}{@{\extracolsep{\fill}}lccc}
\hline
System & Element & Minimum Content (\%) & Maximum Content (\%) \\
\hline
Ag-Pd-Ru & Ag & 10 & 40 \\
         & Pd & 23 & 87 \\
         & Ru & 0  & 45 \\
\hline
Ag-Pd-Pt & Ag & 1  & 70 \\
         & Pd & 0  & 47 \\
         & Pt & 17 & 69 \\
\hline
Ag-Pd-Pt-Ru & Ag & 3  & 39 \\
            & Pd & 0  & 28 \\
            & Pt & 0  & 56 \\
            & Ru & 7  & 67 \\
\hline
\hline
\end{tabular*}
\end{table*}

\begin{table*}
\small
\caption{Comparative metrics of current at 850\,mV across systems and their correlations with elements.}
\label{tbl:current850mVCorrelation}
\begin{tabular*}{\textwidth}{@{\extracolsep{\fill}}llll}
\hline
Metric & Ag-Pd-Ru & Ag-Pd-Pt & Ag-Pd-Pt-Ru \\
\hline
Mean Current (mA) & -0.278 & -0.342 & -0.159 \\
Standard Deviation (mA) & 0.114 & 0.098 & 0.074 \\
Minimum Current (mA) & -0.673 & -0.583 & -0.366 \\
25\% Quantile (mA) & -0.348 & -0.423 & -0.195 \\
Median (mA) & -0.248 & -0.372 & -0.131 \\
75\% Quantile (mA) & -0.189 & -0.271 & -0.110 \\
Maximum Current (mA) & -0.065 & -0.063 & -0.060 \\
Correlation with Ag & +0.766 & +0.587 & +0.440 \\
Correlation with Pd & -0.905 & -0.771 & -0.502 \\
Correlation with Pt & N/A & -0.017 & -0.771 \\
Correlation with Ru & +0.719 & N/A & +0.719 \\
\hline
\end{tabular*}
\end{table*}

\subsubsection{Ag-Pd-Ru system}

\begin{figure}
    \centering
    \includegraphics[width=1\linewidth]{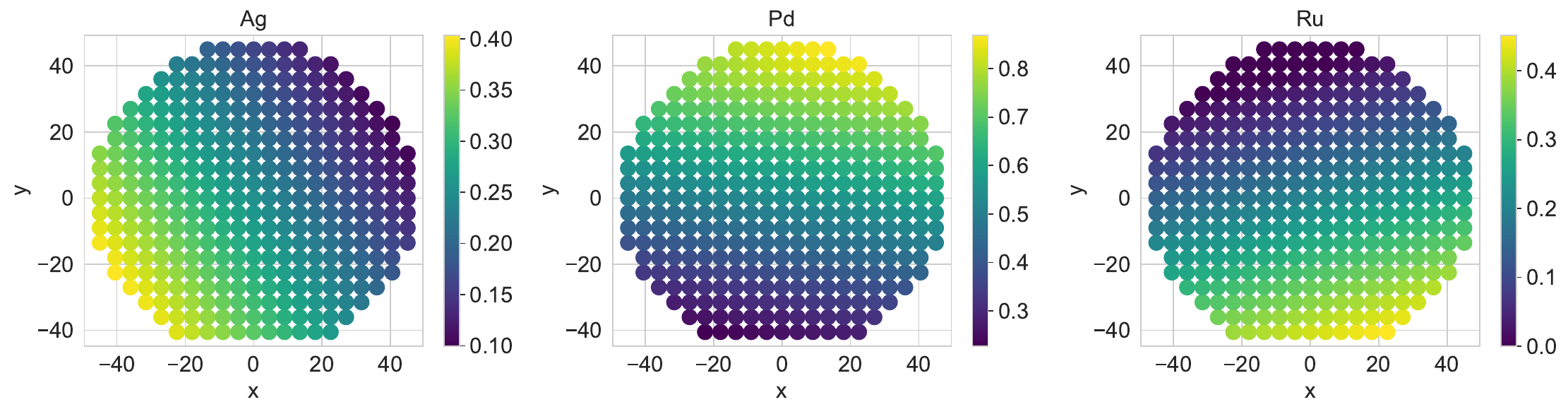}
    \caption{Color-coded plot of compositional gradients in Ag-Pd-Ru system.}
    \label{fig:Ag-Pd-Ru}
\end{figure}

\begin{figure}
    \centering
    \includegraphics[width=0.66\linewidth]{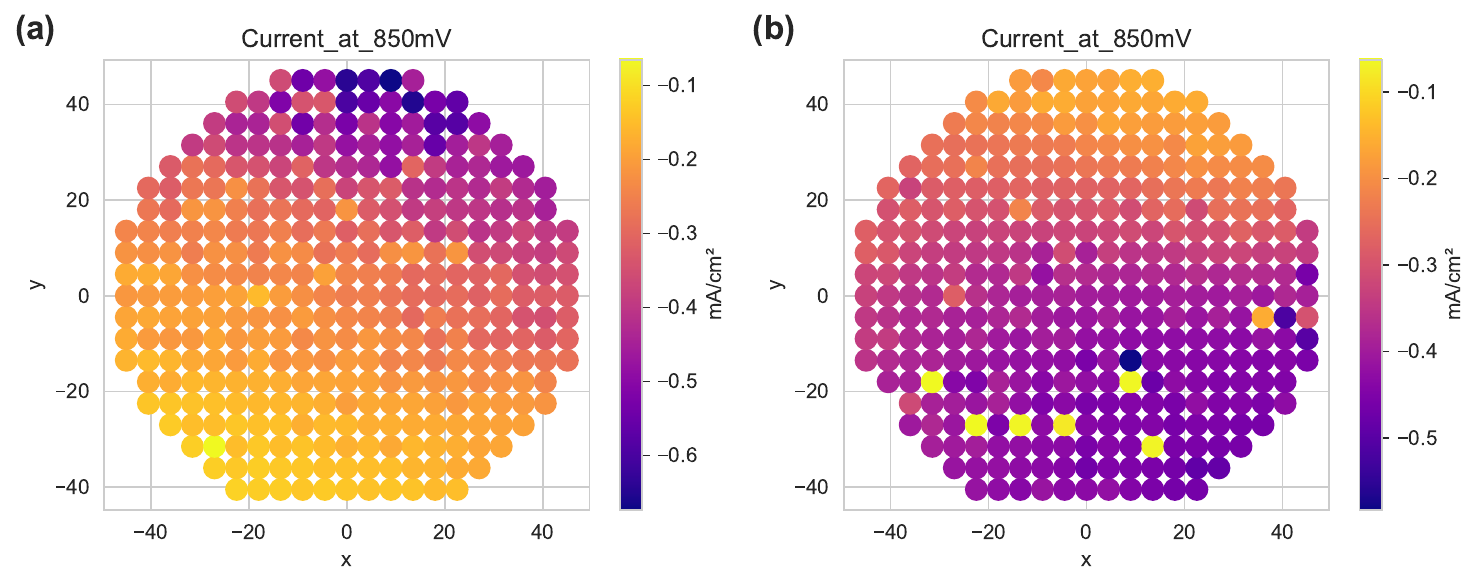}
    \caption{Color-coded plot of current density gradients in: (a) Ag-Pd-Ru system and (b) Ag-Pd-Pt system.}
    \label{fig:ternary_datasets_current}
\end{figure}

The Ag-Pd-Ru system contains a range of element composition, with Pd showing the highest compositional range from 23\% to 87\%, followed by Ru ranging from 0\,\% to 45\,\% and Ag from 10\,\% to 40\,\%.
In terms of electrochemical performance, this system shows a mean current in ORR of -0.278\,mA at 850\,mV. A correlation analysis reveals a significant negative correlation of Pd with electrochemical performance (-0.905), suggesting that higher contents of Pd lead to improved performance (lower current indicated better performance). Conversely, Ru and Ag show positive correlations, +0.719 and +0.766 respectively, indicating that increases in their contents may not favor performance. This suggests that optimizing Pd content while minimizing Ru and Ag could enhance the system's efficiency(Fig.~\ref{fig:Ag-Pd-Ru},Fig.~\ref{fig:ternary_datasets_current} (a)), in line with chemical intuition~\cite{Antolini2009915}.

\subsubsection{Ag-Pd-Pt system}

\begin{figure}
    \centering
    \includegraphics[width=1\linewidth]{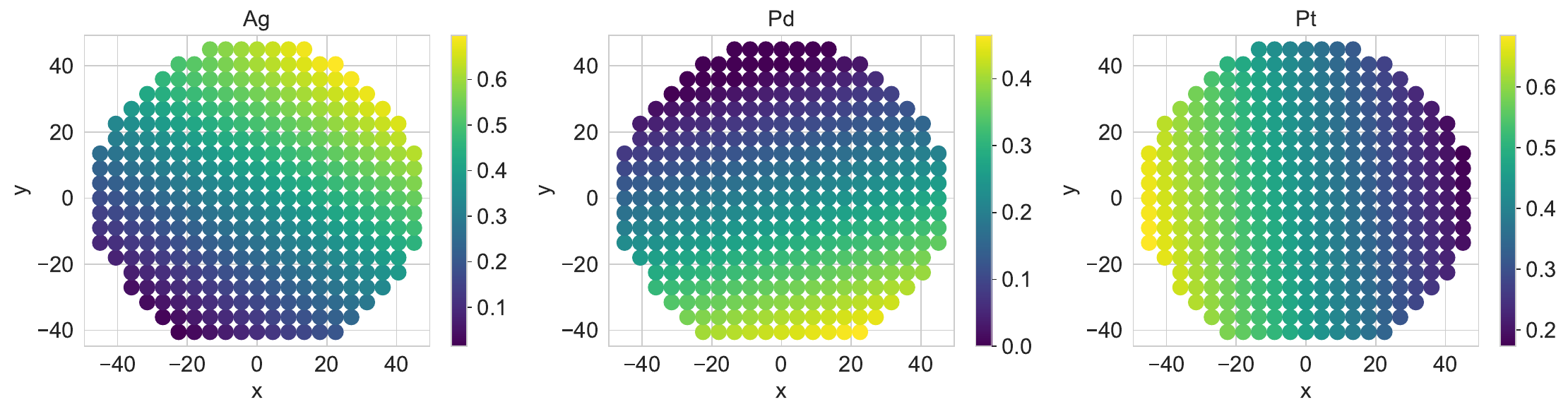}
    \caption{Color-coded plot of compositional gradients in Ag-Pd-Pt dataset}
    \label{fig:Ag-Pd-Pt}
\end{figure}

The Ag-Pd-Pt system exhibits a compositional range with Pd between 0\,\% and 47\,\%, Ag between 1\,\% and 70\,\%, and Pt between 17\,\% and 69\,\%.
The mean current at 850\,mV for the Ag-Pd-Pt system is -0.342\,mA, displaying a slightly better performance compared to the Ag-Pd-Ru system. The correlation analysis shows a strong negative correlation with Pd (-0.771) and a very weak negative correlation with Pt (-0.017), suggesting that Pt's influence on performance is minimal. Ag's positive correlation (+0.587) further implies that, similar to the Ag-Pd-Ru system, increasing Ag content does not benefit the system's performance (Fig.~\ref{fig:Ag-Pd-Pt},Fig.~\ref{fig:ternary_datasets_current} (b)).

\subsubsection{Ag-Pd-Pt-Ru system}

The quaternary system displays a spread of elemental composition with Ru vary from 7\,\% to 67\,\%, Pd from 0\,\% to 28\,\%, Ag from 3\,\% to 39\,\%, and Pt from 0\,\% to 56\,\%. (Table~\ref{tbl:elementalConcentrations}). The performance metrics show a mean current of -0.159\,mA at 850\,mV, which is less negative than the other two systems, suggesting a comparative decrease in performance (Table~\ref{tbl:current850mVCorrelation}).
\begin{figure}
    \centering
    \includegraphics[width=0.75\linewidth]{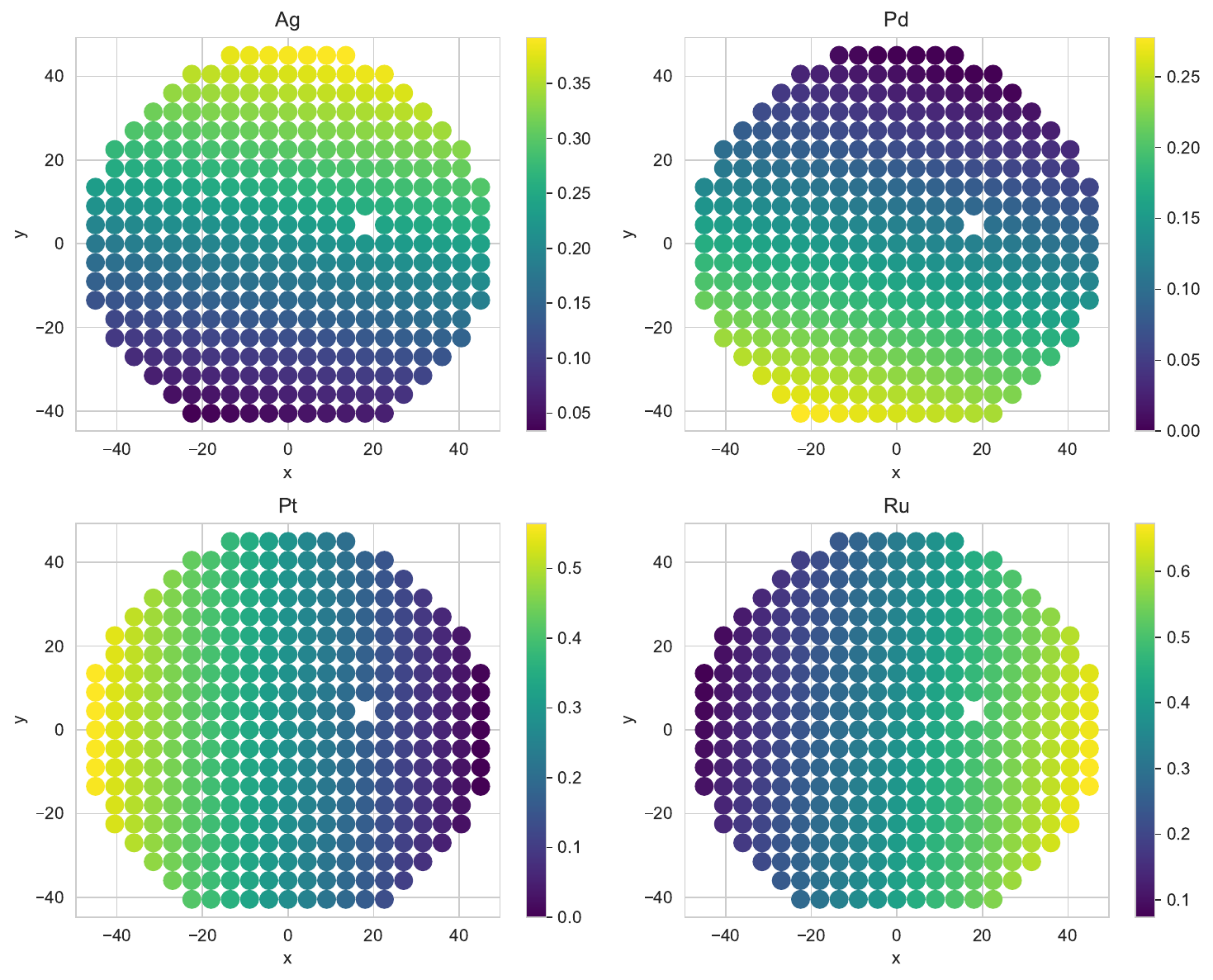}
    \caption{Color-coded plot of compositional gradients in Ag-Pd-Pt-Ru system.}
    \label{fig:Ag-Pd-Pt-Ru_dataset}
\end{figure}
The correlation coefficients present a complex picture. Pd's negative correlation (-0.502) is less pronounced than in the other systems, indicating its diminished influence in the presence of Pt, which shows a strong negative correlation (-0.771) with the current. This suggests that in this system, Pt plays a more critical role in enhancing performance than Pd. Ru and Ag show positive correlations, similar to the Ag-Pd-Ru system, suggesting their less favorable impact on performance.
Overall, the dataset and correlation analysis suggest that the content of Pd and Pt significantly influences the electrochemical performance of these systems, with a higher content of Pd generally leading to better performance, except in the presence of Pt, which can also play a significant role in the Ag-Pd-Pt-Ru system.

\begin{table*}
\small
\caption{Consolidated statistical analysis of actual vs. predicted electrochemical currents across different models.}
\label{tab:prediction_analysis}
\begin{tabular*}{\textwidth}{@{\extracolsep{\fill}}lccccc}
\hline
Metric & Gaussian Process (GP) & GP with Embeddings & Standard Vector Method \\
\hline
Mean (Actual) (mA/cm\(^2\)) & -0.16 & -0.16 & -0.16 \\
Mean (Predicted) (mA/cm\(^2\)) & -0.22 & -0.15 & - \\
Standard Deviation (Actual) (mA/cm\(^2\)) & 0.07 & 0.07 & 0.07 \\
Standard Deviation (Predicted) (mA/cm\(^2\)) & 0.07 & 0.05 & - \\
Minimum (Actual) (mA/cm\(^2\)) & -0.37 & -0.37 & -0.37 \\
Minimum (Predicted) (mA/cm\(^2\)) & -0.35 & -0.07 & - \\
Mean Absolute Error (MAE) & 0.06 & 0.03 & - \\
Root Mean Square Error (RMSE) & 0.07 & 0.04 & - \\
Overall coefficient of determination ($r^2$) & \ 0.08 & \ 0.65 & - \\
Overall Correlation (\(r\)) & 0.85 & 0.83 & 0.79 \\
Correlation (\(r\)) for Current < -0.2 mA/cm\(^2\) & \ 0.63 & \ 0.60 & \ 0.89 \\
\hline
\end{tabular*}
\end{table*}

\subsection{Results of method 1: GP Model with elemental composition}
\begin{figure}
    \centering
    \includegraphics[width=.75\linewidth]{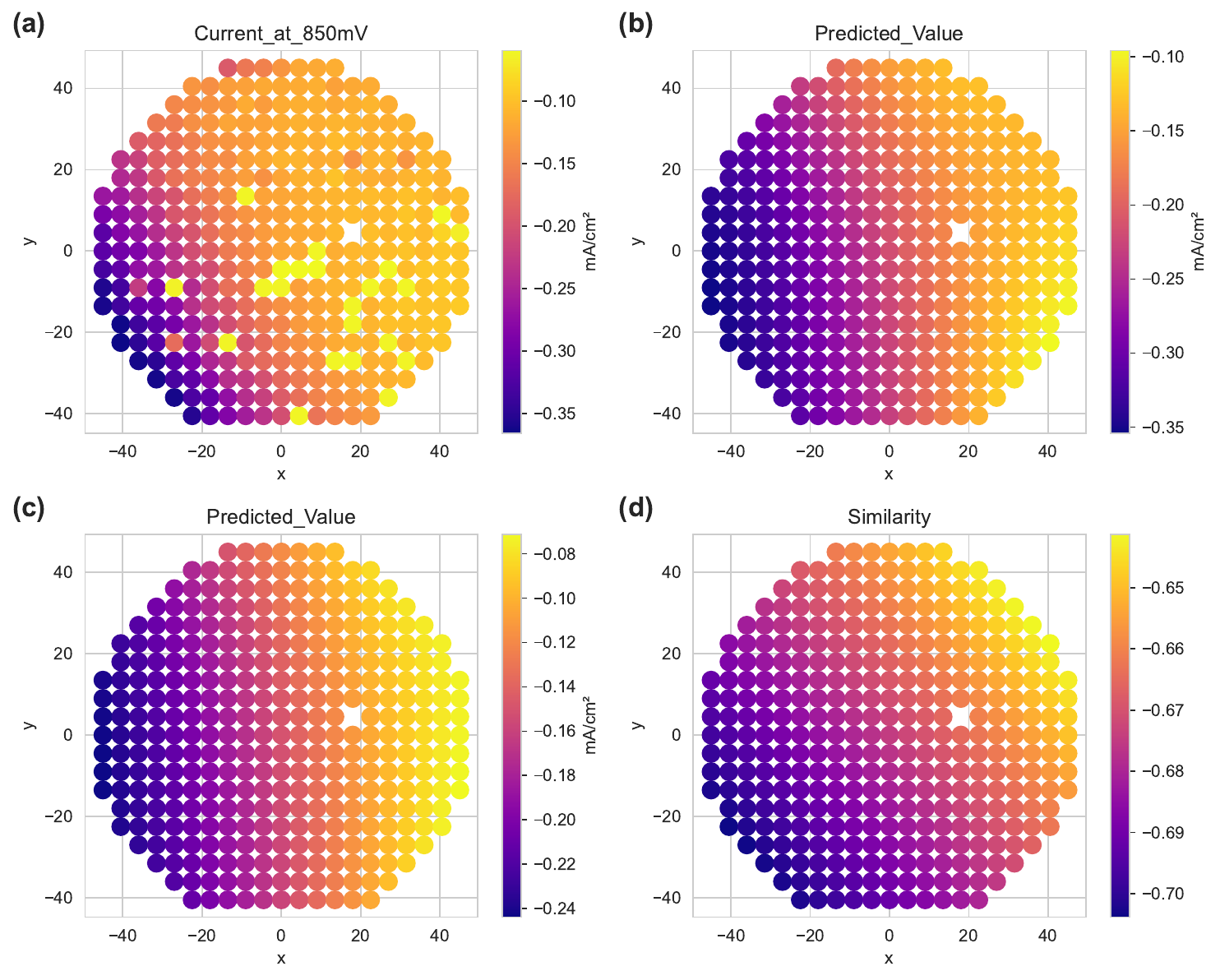}
    \caption{Experimental results of Ag-Pd-Pt-Ru system (a) and prediction results using GP model (b), enhanced GP model with material vectors (c) and standard vector method (d).}
    \label{fig:quaternary_current_with_predictions}
\end{figure}

Table~\ref{tab:prediction_analysis} and Fig.~\ref{fig:quaternary_current_with_predictions} (a,b) present the results of the application of Gaussian Process (GP) based solely on elemental compositions. This approach demonstrates a baseline predictive capability with an overall correlation coefficient (\(r\)) of 0.85 and a coefficient of determination (\(R^2\)) of 0.08. The mean actual electrochemical current was measured at -0.16\,mA/cm\(^2\) with a standard deviation of 0.07\,mA/cm\(^2\). The model's predictions deviated slightly, with a mean predicted current of -0.22\,mA/cm\(^2\) and a comparable standard deviation of 0.07\,mA/cm\(^2\). This method demonstrates a Mean Absolute Error (MAE) of 0.06\,mA/cm\(^2\) and a Root Mean Square Error (RMSE) of 0.07\,mA/cm\(^2\), indicating a moderate level of accuracy in the predictions.

\subsection{Results of method 2: Enhanced GP model with material vectors}
The GP model's performance significantly improved using a word embedding-derived representation of materials as input (Table~\ref{tab:prediction_analysis}, Fig.~\ref{fig:quaternary_current_with_predictions} (a,c)). Most notably, the overall \(R^2\) increases to 0.65, indicating that the model accounts for a much larger proportion of the variance in the data. This suggests a significantly stronger relationship between the predictions and actual measurements when using material vectors.
While the correlation coefficient (\(r\)) slightly decreases to 0.83, the model's ability to capture the general trend of the dataset is markedly improved.
This is evidenced by the mean predicted current of -0.15\,mA/cm\(^2\), which closely matches the actual mean current.
Additionally, with a standard deviation of 0.05\,mA/cm\(^2\), the predictions are more precise compared to the composition-based representation.  Finally, the MAE and RMSE values decreased to 0.03\,mA/cm\(^2\) and 0.04\,mA/cm\(^2\), respectively, further confirming the improved accuracy of the model using material vectors. 

\subsection{Results of method 3: Standard vector method}
\begin{figure}[h!]
    \centering
    \includegraphics[width=0.9\textwidth]{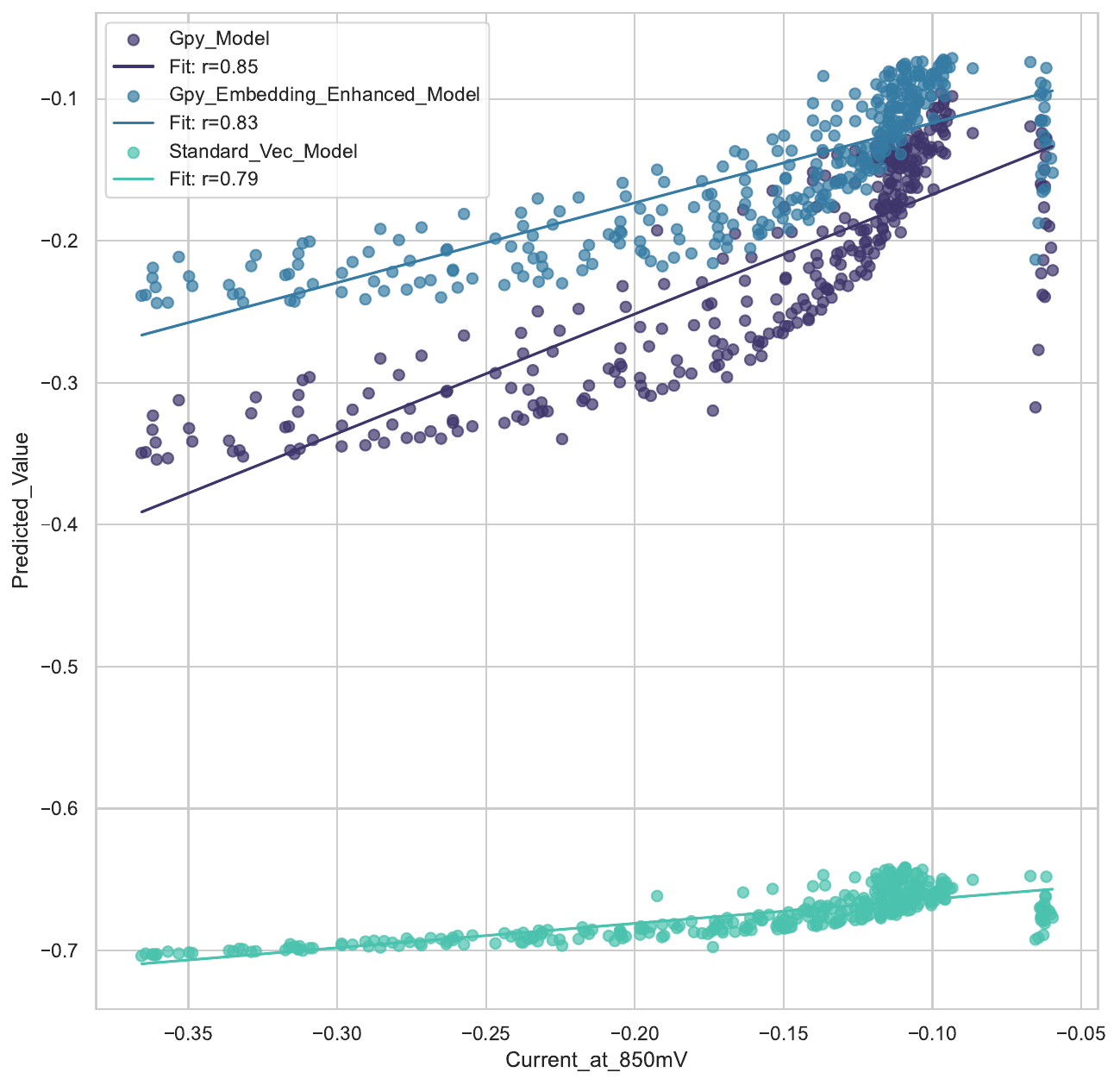}
    \caption{Overall correlation across different systems.}
    \label{esi:cor}
\end{figure}

\begin{figure}[h!]
    \centering
    \includegraphics[width=0.9\textwidth]{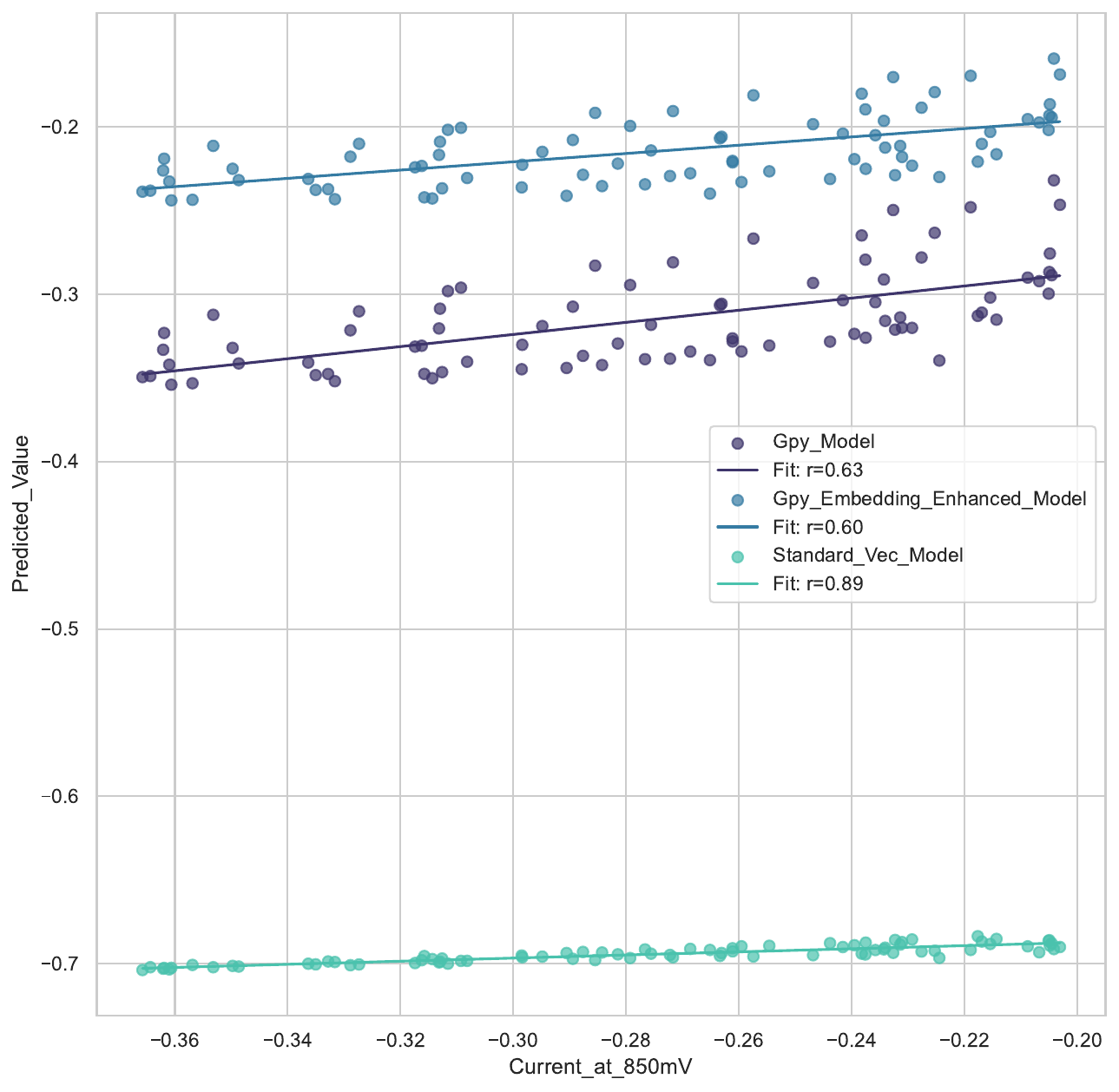}
    \caption{Correlation within filtered region across different systems.}
    \label{esi:cor_f}
\end{figure}

The standard vector approach which uses weighted vector representations of material properties results in very promising improvements of the prediction (Table~\ref{tab:prediction_analysis}, Fig.~\ref{fig:quaternary_current_with_predictions} (a,d), supplementary Fig.~\ref{esi:cor}, Fig.~\ref{esi:cor_f}).
Specific statistical metrics are not provided for this model such as \(R^2\), MAE, or RMSE because we do not predict the current directly but a similarity measure which strongly correlates with the currents at 0.89. This proves a significant correlation with the quaternary system's material performance, particularly in predicting lower electrochemical currents, that is predicting compositions with higher eletrocatalytic performance, with high accuracy which are promising candidates for experimental assessment.

\begin{figure}[ht]
    \centering
    \includegraphics[width=1\linewidth]{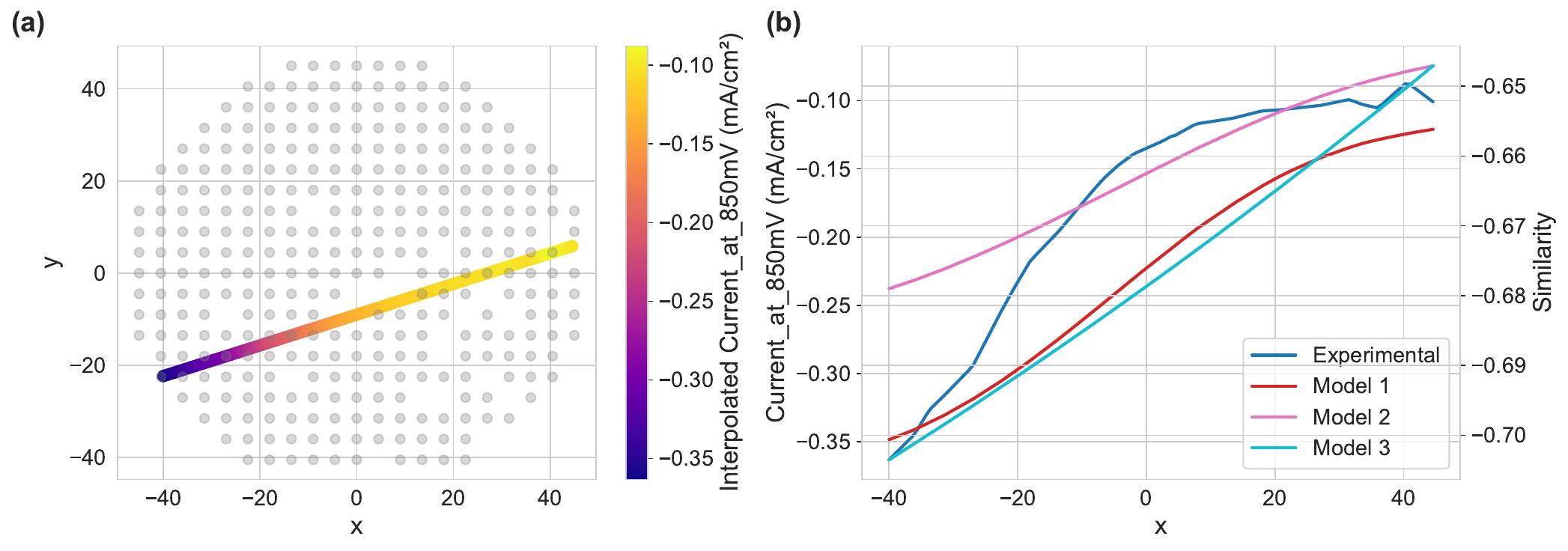}
    \caption{Interpolation results across the whole dataset: (a) Illustration of used line from the maximum and minimum current values with interpolated results, (b) Experimental data and predictions from all models along the direction indicated in (a).}
    \label{fig:whole_region}
\end{figure}

Figure~\ref{fig:whole_region} shows all model predictions in comparison to the experimental data discarding outliers above a threshold of -0.075\,mA/cm\(^2\) along a line across the CSML from the minimum to the maximum of the activity.
The location of the measured data points are shown as gray background markers, the color-coded line represents the continuous interpolation of current values across this direction.
Figure~\ref{fig:whole_region} (b) shows the predictions from the three models along the interpolated measured data.
It is notable that the GP model captures the non-linear behavior of the data more effectively while the standard vector method exhibits noticeable deviations w.r.t. the trend across the CSML.
Nevertheless, the word embedding-based material representation and the standard vector method offer greater flexibility.
Unlike the GP model, which is fixed to the specific dataset, in particular its elements, the other approaches are applicable to other material compositions. Future work will explore non-linear combinations which likely improve the accuracy of the proposed standard vector approach.

\section{Discussion}

\subsection{Interpretation of results}
The outcomes of our study demonstrate that the choice of representation in computational models is critical for prediction performance.
Model 1, GP based on elemental composition provides a reference prediction.
However, its comparatively lower predictive accuracy (\(R^2\) of 0.08) suggest complex (nonlinear) interplay of composition and catalytic performance in the quaternary system, where interactions between elements may not be fully captured using only a compositional representation.

Model 2, the GP model based on word-embedding based representations of materials, shows a significant improvement in predictive accuracy (\(R^2\) of 0.65). 
We attribute this improvement to the latent knowledge captured through word embedding representations of the compositions.
It demonstrates that the complex interactions between materials beyond elemental composition can be captured in representations and effectively used for prediction.

Model 3, the standard vector approach, further exploits relationships of word embeddings by not directly predicting performance but instead focusing on the optimization of a similarity measure between materials vector representation and a standard vector based on known correlations of certain terms with electrocatalytic performance and experimental data from the two ternary systems.
The high correlation (\(r\)) of 0.89 for specific conditions indicates a success, emphasizing the method's capability to identify potential high-performing materials within a defined search space.
Our approach highlights the potential of using latent knowledge from scientific literature about materials and their relationships and represents a new approach for the representation of materials in combination with experimentally measured data.

\subsection{Comparison with existing literature}

Our findings resonate with and extend existing research in materials science, particularly the use of machine learning and vector-based representations for materials prediction~\cite{Huang2023,Xiong2020}. Several studies have demonstrated the potential of machine learning models, especially those incorporating innovative data representations, to outperform traditional computational methods~\cite{Tian2022}.
Our work aligns with these findings, showcasing the effectiveness of material vectors for capturing complex interactions.
However, we introduce a unique focus on similarity measures combined with word embedding-derived representations of materials, a less explored approach within materials property predictions.

\subsection{Advantages of the proposed methods}
Word embedding-based representations are directly combined with experimental data to predict unknown, more complex composition-property spaces.
By using latent knowledge encoded in word embeddings we counterbalance data scarcity typically prevalent in experimental discovery campaigns, thereby accelerating the discovery process.

Our standard vector approach introduces a novel approach by focusing on `similarity' rather than direct prediction.
Our method's success in identifying high-performing materials based on their similarity to an optimized standard vector highlights based on experimental data is a tool for material selection and discovery, especially in systems where direct performance data may is scarce or hard to predict because of yet-unknown correlations.
In our approach, we combine reliable but expensive-to-obtain experimental data with the fuzzy but cheap-to-obtain correlations in word embeddings.
Our `standard vector' can be viewed as a electrocatalysis-specific sequence of materials features\cite{Rajan2015} for specific materials systems and is particularly useful in scenarios where data is scarce.

\subsection{Limitations and challenges}
While our methods demonstrate significant advancements, they are not without limitations.
For one, the word embeddings depend on the literature data set. We have restricted ourselves to literature with open access licenses.
More text data, e.g. from copyright-protected material, could in principle improve word embeddings.
The past and current publishing routes, however, restrict usage of the knowledge in literature without special agreements with publishers.
Second, we rely on comprehensive and accurately labeled (ideally experimental) datasets for training the models and finding the `standard vector'.
This, in general poses a challenge, particularly in material science, where experimental data can be scarce, incomplete, or inconsistent.
Additionally, the complexity of the models, especially the standard vector approach, may introduce difficulties in interpretation and implementation, potentially limiting their accessibility for broader application.

Future research will focus on addressing these limitations, possibly through the development of more robust models that can handle even more sparse or noisy data, and the exploration of methods to simplify model interpretations without sacrificing prediction accuracy.

\subsection{Implications for future research}
Our study highlights the usefulness of material vectors and similarity measures for predicting material performance, paving the way for advancements in materials prediction for under-explored compositional spaces where partial high-quality data already exists. Here are specific directions for future research:

\textbf{Integration with experimental approaches:} Combining these computational methods with targeted experimental validation can lead to iteratively more refined models and accelerated materials discovery.
Experiments can verify predictions, identify regions where models need improvement, and provide new data to further enhance predictive power to include elements for predictions of different properties.

\textbf{Hybrid models:} Combining our methods with other predictive techniques like ab initio simulations or machine learning algorithms\cite{Clausen2024} could create more robust hybrid models.
These models could leverage the strengths of different approaches, potentially addressing shortcomings and enhancing predictive accuracy across varied, multimodal datasets.

\textbf{Complex material systems:} The success shown in this study encourages applying these methods to other properties of complex material systems.
These could include structural, energy storage, magnetic properties, etc., i.e. any system where properties are mainly a function of composition and not of microstructure.
In contrast to composition-based models as presented here, the word embedding-derived representations make arbitrary choices of elemental combinations seamless.
We expect that the near future will allow to use more experimental data for refinement of `standard vectors'. 
Provided more reliable data for specific composition-property relationships is be available, `standard vectors' for specific use cases could be defined as references against which new compositions could be assessed.
New compositions could then be judged w.r.t. (theoretical) suitability be useful for a specific use case.
If several such standard vectors can be defined, new compositions could be assessed for their suitability for multi-functional purposes.

\section{Conclusions}
Our study has successfully demonstrated the potential of machine learning and vector analysis techniques in predicting materials performance in ternary and quaternary compositionally complex solid solutions based on parameter-free Gaussian Process (GP) and literature-derived materials representations.
The use of a GP model with elemental composition established a baseline for predictive accuracy, achieving a coefficient of determination value ($r^2$) of 0.08.
An improved version of the GP model based on material vectors as representations for the composition derived from literature mining marks a significant improvement, with an improved $r^2$ value of 0.65.
However, the most notable advancement was achieved with our proposed similarity vector approach.
This method, which relies on the construction and optimization of property vectors, demonstrates a remarkable correlation with experimental outcomes, evidenced by a correlation value of 0.89. The superior performance underscores the potential of word embedding-based methods to leverage knowledge and material correlations from existing literature.

\section*{Author Contributions}
\textbf{Lei Zhang:} Conceptualization, Methodology, Software, Validation, Formal analysis, Investigation, Data Curation, Writing - Original Draft, Visualization, Experimentation, Funding acquisition.\\
\textbf{Lars Banko:} Synthesis and characterization of materials libraries, pre-processing of the dataset, Editing\\
\textbf{Wolfgang Schuhmann}: Electrochemical experimentation, Supervision, Writing - Review.\\
\textbf{Alfred Ludwig:} Conceptualization, Writing - Review \& Editing.\\
\textbf{Markus Stricker:} Conceptualization, Resources, Supervision, Writing - Review \& Editing.
\section*{Conflicts of interest}
There are no conflicts to declare.

\section*{Acknowledgements}
The contribution of Dr. Olga Krysiak for performing the SDC measurements is acknowledged.
Lei Zhang and Markus Striker gratefully acknowledge the financial support provided by the China Scholarship Council (CSC, CSC number: 202208360048), which was instrumental in facilitating this research. 
All authors acknowledge funding by the Deutsche Forschungsgemeinschaft (DFG, German Research Foundation) – CRC 1625, project number 506711657, subprojects INF, A05, A01, C01.

\bibliographystyle{unsrt}  
\bibliography{references}  

\newpage
\section*{Supplementary Material}

\end{document}